\documentclass{article}
\usepackage{amssymb}
\usepackage[all]{xy}
\usepackage{authblk}

\def\R{{\mathbb R}}

\newtheorem{theorem}{Theorem}

\begin{document}

\title{Sustained oscillations in the MAP kinase cascade} 
\author[1]{Juliette Hell}
\author[2]{ Alan D. Rendall}
\affil[1]{Freie Universit\"at Berlin, { \tt jhell@zedat.fu-berlin.de}.}
\affil[2]{Johannes Gutenberg-Universit\"at Mainz, { \tt rendall@uni-mainz.de}.}
\date{}

\date{}

\maketitle
\abstract{The MAP kinase cascade is a network of enzymatic reactions arranged in layers. In each layer occurs a multiple futile cycle of phosphorylations. The fully phosphorylated substrate then serves as an enzyme for the layer below. This papers focusses on the existence of parameters for which Hopf bifurcations occur and generate periodic orbits. Furthermore it is explained how geometric singular perturbation theory allows to generalize results from simple models to more complex ones.}

\section{Introduction}

The MAP kinase cascade (mitogen-activated protein kinase cascade) is a pattern
of chemical reactions encountered frequently in cell biology. The actual 
substances involved in the reactions vary from one example to another but 
certain features are always present. There are three proteins, generically
denoted by MAPK, MAPKK and MAPKKK. For brevity we mostly use the notations 
$K$, $KK$ and $KKK$ for these.
 A phosphate group can be attached to $KKK$ at a 
particular site and this causes it to become activated. It then catalyses the 
addition of phosphate groups to $KK$ at two sites. An enzyme which catalyses a 
phosphorylation in this way is called a kinase and MAPKKK stands for MAPKK 
kinase. Similarly when $KK$ has been phosphorylated at both sites it becomes 
activated and catalyses the addition of phosphate groups to $K$ at two sites. 
There are also other enzymes (phosphatases) which remove phosphate groups from 
each of the three proteins. We refer to one of the proteins $K$, $KK$ or $KKK$ 
together with its phosphorylated forms and the reactions interconverting these 
forms as a layer of the cascade. We think of the layers corresponding to 
$KKK$, $KK$ and $K$ as being arranged from top to bottom and refer to them as 
the first, second and third layers. 
Since substances in one layer directly 
influence the reactions in the next layer information flows from the top to 
the bottom in this picture and that is the reason it is called a cascade.
The MAPK cascade can be represented in the following way:
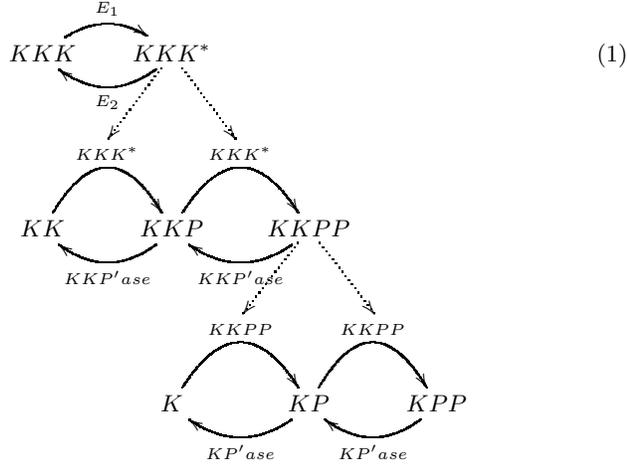
\begin{equation}\label{cascade}
\xymatrix@C=0.5em{
 KKK\ar@/^1pc/[rr]^{E_1}&&KKK^* \ar@/^1pc/[ll]^{E_2}   \ar@{.>}[dl]  \ar@{.>}[dr]&&&&\\
 &&&&&&\\
 KK\ar@/^2pc/[rr]^{KKK^*} &&KKP\ar@/^2pc/[rr]^{KKK^*} \ar@/^1pc/[ll]^{KKP'ase}  &&KKPP\ar@/^1pc/[ll]^{KKP'ase} \ar@{.>}[dl]  \ar@{.>}[dr] &&\\
 &&&&&&\\
 &&K\ar@/^2pc/[rr]^{KKPP}&&KP\ar@/^2pc/[rr]^{KKPP} \ar@/^1pc/[ll]^{KP'ase}&&KPP\ar@/^1pc/[ll]^{KP'ase}  
}
\end{equation}
In this diagram, an enzymatic reaction $\xymatrix{X \ar[r]^{E} &Y}$ represents the chemical reactions $X+E \rightleftarrows X\cdot E \rightarrow Y+E$, where $X\cdot E$ is a complex of substrate $X$ and enzyme $E$. 
In 
fact information can flow upwards through the cascade in a more indirect sense 
and this is described in more detail below. This backward flow has an 
essential influence on the dynamics of the system.

If each of the reactions is modelled by a standard irreversible 
Michaelis-Menten scheme (see for instance \cite{murray89}) composed of three 
elementary reactions with substrate, enzyme and substrate-enzyme complex and 
the elementary reactions are given mass action kinetics a system of ordinary 
differential equations arises as a model for the MAPK cascade. This will be 
referred to as the Huang-Ferrell model since it was introduced by those authors
in \cite{huang96}. In that paper they studied mathematical properties of these 
equations and also compared solutions of the equations with the results of the 
experiments they had done on a MAPK cascade in extracts of oocytes (immature 
egg cells) of the frog {\it Xenopus}. The full system consists of 22 equations 
(for 8 substrates, 4 enzymes which do not occur as substrates and 10 complexes)
and depends on 30 parameters (reaction constants). The procedure in 
\cite{huang96} was to fix all but one of the parameters and numerically 
determine stationary solutions of the system for different values of the 
remaining parameter. Then the value of the concentration of one of the 
substances at the stationary solution was studied as a function of the chosen 
parameter. This function exhibits the property of ultrasensitivity where the 
output is a sigmoid function of the input.

The possibility of carrying out this procedure is dependent on the fact that 
there exist a stationary solution for each value of the parameters. Moreover
the results will only be unambiguous if there is only one such solution for 
fixed values of the reaction constants and the total amounts of the three 
proteins and the four other enzymes. This issue is not addressed in 
\cite{huang96}. That the answer to the uniqueness question is not obvious
is made clear by the results of \cite{markevich04}. That paper was concerned 
with a system which can be thought of as a single layer of the MAPK cascade with
two phosphorylation steps. The ODE system in this case is known as the
dual futile cycle \cite{wangsontag08a}. The numerical and heuristic work in 
\cite{markevich04} indicated that this system exhibits bistability, i.e. that 
there are parameter values for which there exists more than one stable 
stationary solution. A rigorous proof that this is the case was given in 
\cite{hell14}.

A priori it is not ruled out that solutions of the Huang-Ferrell model might
exhibit more complicated long-time behaviour than just convergence to a 
stationary solution. In fact, numerical and heuristic work in \cite{qiao07}
indicates that periodic solutions exist. There is also evidence suggesting
that chaotic behaviour may occur \cite{zumsande10}. According to the 
investigations of \cite{qiao07} periodic solutions already occur in the system
corresponding to the first two layers of the MAPK cascade. We refer to this 
system as the truncated Huang-Ferrell model. In what follows we will prove
that for both the truncated and full Huang-Ferrell models there exist 
parameters for which there are periodic solutions. This means that the protein 
concentrations being modelled undergo sustained oscillations.  

Following the terminology of \cite{grimbs11} we refer to the Huang-Ferrell 
model or an analogous system for a subset of the layers of the cascade as the 
MM-MA system (for Michaelis-Menten via mass action). Under suitable 
circumstances it is possible to derive a smaller system, the MM system (for 
Michaelis-Menten), via a quasistationary approximation. This is relatively 
simple to do for the dual futile cycle and the dynamics of the MM system in 
that case was studied in \cite{ortega06}. In that paper the authors found 
that bistability is already present in the MM system. The method used in 
\cite{hell14}, which will be generalized
here, is to first prove bistability for the MM system and then use the fact
that the MM system is a limit of the MM-MA system in a suitable sense to 
obtain the corresponding result for the MM-MA system. For this we used that 
the stationary solutions are hyperbolic. The technique applied to treat the 
singular limit is geometric singular perturbation theory (GSPT) 
\cite{fenichel79}. In \cite{hell14} an MM limit was 
defined for the truncated Huang-Ferrell model. The definition was inspired
by ideas in \cite{ventura08} and \cite{ventura13}. In these papers it 
was pointed out that the phenomenenon of sequestration can lead to a flow
of information from layers further down in the cascade to higher levels. 
For instance, if a lot of $KKK^*$, the activated form of $KKK$, is bound
to its substrates $KK$ and the phosphorylated form $KKP$ then not much of it 
is available as a substrate for the phosphatase which would otherwise convert 
it back to the inactivated form $KKK$. Related ideas are discussed in 
\cite{delvecchio08}, where this effect is called retroactivity. It was 
mentioned in \cite{hell14} that if it could be proved that this MM system 
admits a hyperbolic periodic solution then the truncated Huang-Ferrell system 
would also admit a periodic solution. It turned out to be difficult to obtain a 
proof of hyperbolicity and thus we will use a slightly different 
strategy in what follows. The basic idea is nevertheless still to first prove
a result for the MM system and then lift some of the structure found to the 
MM-MA system.

The paper is organized as follows. Section 2 introduces some basic notation
and terminology and Section 3 proceeds to show that the MM system derived
from the truncated Huang-Ferrell model admits a Hopf bifurcation for certain 
values of the parameters. It is then shown that it can be concluded that the 
MM-MA system also admits a Hopf bifurcation. It follows from the basic theorem 
of Hopf (cf. \cite{hale69}) that there are parameters for which the MM-MA 
system admits periodic solutions. These results are summed up in Theorem \ref{theorem1}.
Section 4 shows that the bistability in the dual futile cycle implies the 
presence of bistability in the truncated Huang-Ferrell model. In Section 5 the 
arguments of Section 3 are extended so as to prove Theorem \ref{theorem2} which asserts the 
existence of periodic solutions for the full Huang-Ferrell model. In 
particular this involves the use of an MM system for the full MAPK cascade. 
Section 6 presents results on some variants of the models coming from the 
MAPK cascade.
 For a cascade consisting of two single phosphorylation loops
it is proved that all solutions of the MM system converge to the same 
stationary solution. 
In particular there are no sustained oscillations for 
that model. 
For a cascade consisting of a layer with two phosphorylations 
above a layer with one phosphorylation it is not clear whether sustained oscillations can occur. It is shown how this configuration can arise by modelling a system considered in \cite{prabakaran14}, thus motivating further study of this question.  
The final section indicates some directions 
in which the results of this paper might be extended. Some information on 
geometric singular perturbation theory is collected in an appendix.

\section{The basic equations}\label{basic}

In what follows we essentially adopt the notations of Huang and Ferrell 
\cite{huang96} for the full cascade and then specialize them to the truncated 
system. The unknowns in the system are the concentrations 
\begin{eqnarray}\label{substances}
&&[KKK], [KKK^*], [E_1], [E_2], [KKK\cdot E_1], [KKK^*\cdot E_2],\nonumber\\ 
&&[KK], [KKP], [KKPP], [KK P'ase],\nonumber\\ 
&&[KK\cdot KKK^*], [KKP\cdot KKK^*],
[KKP\cdot KK P'ase], [KKPP\cdot KK P'ase],\nonumber\\
&&[K], [KP], [KPP], [K P'ase]\nonumber\\
&&[K\cdot KKPP], [KP\cdot KKPP], [KP\cdot K P'ase], [KPP\cdot K P'ase]  . 
\end{eqnarray}
The square brackets around a symbol indicate the concentration of
the substance denoted by that symbol. Here $E_1$ and $E_2$ are the kinase and 
the phosphatase in the first layer while $KK P'ase$ and $K P'ase$ are the 
phosphatases in the second and third layers. $KKK^*$ is the activated (i.e.
phosphorylated) form of $KKK$. A $P$ occurring in the name of a protein 
indicates a phosphate group. For instance $KKPP$ is the doubly phosphorylated 
form of $KK$. Finally, the substrate-enzyme complexes are denoted by the 
symbols for the two substances separated by a centred dot. The full set of 22 
evolution equations for these quantities will not be reproduced here. They can 
be found in \cite{huang96}. More precisely, 18 of them can be found there and 
the other four, those for the concentrations of the free enzymes, are easily 
derived from those. There are seven conserved quantities, three corresponding 
to the total amounts of the three substrate proteins in all phosphorylation 
states and four corresponding to the total amounts of the four enzymes. These 
conservation laws can be substituted into the evolution equations so as to 
reduce their number. In \cite{huang96} the four conserved quantities 
associated to the enzymes are used in this way while the others are not. Thus 
18 evolution equations remain. The reaction constants for the elementary 
reactions in the Michaelis-Menten description are denoted by $\tilde a_i$  for 
the formation of the complex, $k_i$ for the liberation of product and 
$\tilde d_i$ for the release of the substrate from the complex. These 
notations are as in \cite{huang96} except that $a_i$ and $d_i$ have been 
replaced by $\tilde a_i$ and $\tilde d_i$ since $a_i$ and $d_i$ will be used 
for a different purpose later. All reaction constants are assumed positive
and since the unknowns in the equations are concentrations the solutions of
interest are those which are positive, i.e. all their components are positive. 

The truncated Huang-Ferrell model is obtained by setting the concentrations
of all the unknowns in the full model involving the substance $K$ and that of
the enzyme $KP'ase$ (i.e. the quantities in the last two lines of 
(\ref{substances})) to zero and discarding the evolution equations for those 
quantities. For the rest of this section we concentrate on the truncated 
model. This serves, in particular, to introduce some of the main techniques of
the paper in a context simpler than that of the full cascade. The results for 
the truncated cascade also constitute a central component of the proof of the 
theorems for the full cascade. As explained in \cite{hell14} the passage to 
the limiting MM system can be carried out by scaling the variables in a certain 
way with powers of a parameter $\epsilon$ and then letting $\epsilon$ tend to 
zero. First it is important to 
introduce a new variable to replace $[KKK^*]$. This is
\begin{equation}
\overline {KKK}=[KKK^*]+[KK\cdot KKK^*]+[KKP\cdot KKK^*].
\end{equation}
When the Huang-Ferrell model is written in terms of $\overline {KKK}$ the 
equations of the truncated model are a subset of the equations for the full 
model. The scaling is done as follows. The quantities in the evolution 
equations are replaced by new quantities defined in the following way. The 
concentrations of compounds not containing 
$KKK$ or the enzymes $E_i$ or $KKP'ase$ are not rescaled. These are the first
three quantities in the second line of (\ref{substances}). For the 
concentrations of compounds which contain $KKK$ or $KKP'ase$ but not the $E_i$ 
the new quantity is $\epsilon^{-1}$ times the old one. These are the first
two quantities in the first line, the last quantity in the second line and all 
the quantities in the third line of (\ref{substances}). For the quantities 
involving $E_1$ and $E_2$ the new quantity is $\epsilon^{-2}$ times the old 
one. These are the last four quantities in the first line of 
(\ref{substances}). The reaction constants $\tilde a_1$ and $\tilde a_2$ 
are multiplied by $\epsilon$ to 
get new quantities. All the other reaction constants are left unchanged. In 
addition a new time coordinate is introduced as $\epsilon$ times the old one. 
To avoid complicating the notation the rescaled quantities will be denoted in 
the same way as the original ones. There results the following system, which 
was given in a different notation in \cite{hell14}.
\begin{eqnarray}
&&\frac{d}{dt}(\overline{KKK})=-\tilde a_2[KKK^*][E_2]+\tilde d_2[KKK^*\cdot E_2]
+k_1[KKK\cdot E_1],\label{tev1}\\
&&\frac{d}{dt}[KK]=-\tilde a_3[KK][KKK^*]+\tilde d_3[KK\cdot KKK^*]
+k_4[KKP\cdot KK P'ase],\label{tev2}\\
&&\frac{d}{dt}[KKPP]=-\tilde a_6[KKPP][KK P'ase]+\tilde d_6[KKPP\cdot KK P'ase]
+k_5[KKP\cdot KKK^*],\label{tev3}\\
&&\epsilon\frac{d [KKK\cdot E_1]}{dt}=\tilde a_1[KKK][E_1]
-(\tilde d_1+k_1)[KKK\cdot E_1],\label{tev4}\\
&&\epsilon\frac{d [KKK^*\cdot E_2]}{dt}=\tilde a_2[KKK^*][E_2]-(\tilde d_2+k_2)
[KKK^*\cdot E_2],\label{tev5}\\
&&\epsilon\frac{d [KK\cdot KKK^*]}{dt}=\tilde a_3[KK][KKK^*]-(\tilde d_3+k_3)
[KK\cdot KKK^*],\label{tev6}\\
&&\epsilon\frac{d [KKP\cdot KKK^*]}{dt}=\tilde a_5[KKP][KKK^*]-(\tilde d_5+k_5)
[KKP\cdot KKK^*],\label{tev7}\\
&&\epsilon\frac{d [KKP\cdot KK P'ase]}{dt}=\tilde a_4[KKP][KK P'ase]
-(\tilde d_4+k_4)[KKP\cdot KK P'ase],\label{tev8}\\
&&\epsilon\frac{d [KKPP\cdot KK P'ase]}{dt}=\tilde a_6[KKPP][KK P'ase]
-(\tilde d_6+k_6)[KKPP\cdot KK P'ase].\label{tev9}
\end{eqnarray}
These equations correspond to a subset of the equations given in \cite{huang96}.
To get a closed system we must express some of the quantities on the right hand
side of the equations in terms of those on the left hand side using the 
definition of $\overline{KKK}$ and the conservation laws. The 
necessary equations are
\begin{eqnarray}
&&[KKK^*]=\overline{KKK}-[KK\cdot KKK^*]-[KKP\cdot KKK^*],\\
&&[KKK]=KKK_{\rm tot}-\overline{KKK}+O(\epsilon),\\
&&[KKP]=KK_{\rm tot}-[KK]-[KKPP]+O(\epsilon),\\
&&[E_1]=E_{1,{\rm tot}}-[KKK\cdot E_1],\\
&&[E_2]=E_{2,{\rm tot}}-[KKK^*\cdot E_2],\\
&&[KK P'ase]=(KK P'ase)_{\rm tot}-[KKP\cdot KK P'ase]
-[KKPP\cdot KK P'ase].
\end{eqnarray}
This system of equations extends smoothly to $\epsilon=0$. The terms written
as $O(\epsilon)$ are sums of concentrations of complexes. They are not 
written explicitly since they vanish for $\epsilon=0$ and thus make no
contribution to the limiting equations. When $\epsilon=0$ the equations 
(\ref{tev4})-(\ref{tev9}) become algebraic equations. Substituting 
these into the other evolution equations gives a set of three evolution 
equations, the MM system, which will now be computed. First the concentrations
of the complexes can be expressed in terms of those of the substrates and the
free enzymes.
\begin{eqnarray}
&&[KKK\cdot E_1]=\frac{\tilde a_1[KKK][E_1]}{\tilde d_1+k_1},\\
&&[KKK^*\cdot E_2]=\frac{\tilde a_2[KKK^*][E_2]}{\tilde d_2+k_2},\\
&&[KK\cdot KKK^*]=\frac{\tilde a_3[KK][KKK^*]}{\tilde d_3+k_3},\\
&&[KKP\cdot KKK^*]=\frac{\tilde a_5[KKP][KKK^*]}{\tilde d_5+k_5},\\
&&[KKP\cdot KK P'ase]=\frac{\tilde a_4[KKP][KK P'ase]}{\tilde d_4+k_4},\\
&&[KKPP\cdot KK P'ase]=\frac{\tilde a_6[KKPP][KK P'ase]}{\tilde d_6+k_6}.
\end{eqnarray} 
It is convenient to introduce the Michaelis constants 
$K_{m,i}=\frac{\tilde d_i+k_i}{\tilde a_i}$. Now the concentrations of the free 
enzymes can be expressed in terms of the total concentrations of the enzymes. 
Note that
\begin{eqnarray}
&&E_{1,{\rm tot}}=[E_1](1+K_{m,1}^{-1}[KKK]),\\
&&E_{2,{\rm tot}}=[E_2](1+K_{m,2}^{-1}[KKK^*]),\\
&&\overline{KKK}=[KKK^*](1+K_{m,3}^{-1}[KK]+K_{m,5}^{-1}[KKP]),\\
&&(KK P'ase)_{\rm tot}=[KK P'ase](1+K_{m,4}^{-1}[KKP]+K_{m,6}^{-1}[KKPP]).
\end{eqnarray}
Substituting this into the equations 
\begin{eqnarray}
&&\frac{d}{dt}(\overline{KKK})=-\frac{k_2}{K_{m,2}}[KKK^*][E_2]
+\frac{k_1}{K_{m,1}}[KKK][E_1],\label{inter1}\\
&&\frac{d}{dt}[KK]=-\frac{k_3}{K_{m,3}}[KK][KKK^*]
+\frac{k_4}{K_{m,4}}[KKP][KK P'ase],\label{inter2}\\
&&\frac{d}{dt}[KKPP]=-\frac{k_6}{K_{m,6}}[KKPP][KK P'ase]
+\frac{k_5}{K_{m,5}}[KKP][KKK^*]\label{inter3}
\end{eqnarray}   
gives
\begin{eqnarray}
&&\frac{d}{dt}(\overline{KKK})=
\frac{k_1K_{m,1}^{-1}E_{1,{\rm tot}}[KKK]}{1+K_{m,1}^{-1}[KKK]}\nonumber\\
&&-\frac{k_2K_{m,2}^{-1}E_{2,{\rm tot}}\overline{KKK}}
{1+K_{m,2}^{-1}\overline{KKK}+K_{m,3}^{-1}[KK]+K_{m,5}^{-1}[KKP]},\label{Mmm1}\\
&&\frac{d}{dt}[KK]=-\frac{k_3K_{m,3}^{-1}\overline{KKK}[KK]}
{1+K_{m,3}^{-1}[KK]+K_{m,5}^{-1}[KKP]}\nonumber\\
&&+\frac{k_4K_{m,4}^{-1}(KK P'ase)_{\rm tot}[KKP]}
{1+K_{m,4}^{-1}[KKP]+K_{m,6}^{-1}[KKPP]},\label{Mmm2}\\
&&\frac{d}{dt}[KKPP]=+\frac{k_5K_{m,5}^{-1}\overline{KKK}[KKP]}
{1+K_{m,3}^{-1}[KK]+K_{m,5}^{-1}[KKP]}\nonumber\\
&&-\frac{k_6K_{m,6}^{-1}(KK P'ase)_{\rm tot}[KKPP]}
{1+K_{m,4}^{-1}[KKP]+K_{m,6}^{-1}[KKPP]}.\label{Mmm3}
\end{eqnarray}
Equations (\ref{Mmm1})-(\ref{Mmm3}) are the Michaelis-Menten system. For 
convenience we introduce the notations $a_i=k_{i+2}K_{m,i+2}^{-1}$ for 
$i=1,3$,  $a_i=k_{i+2}K_{m,i+2}^{-1}(KK P'ase)_{\rm tot}$ for $i=2,4$,
$c_i=k_iK_{m,i}^{-1}E_{i,{\rm tot}}$ for $i=1,2$ and $d_i=K_{m,i}^{-1}$ for
$i=1,2$. Furthermore we replace $[KKP]$, $[KKK]$ by their expressions in the chosen phase variables $\overline{KKK}$, $[KK]$, $[KKPP]$and the parameters of the system. 

Assume that the coefficients $K_{m,i}^{-1}$ are equal to a common 
quantity $b_1$ for $3\le i\le 6$. For any choice of the parameters $a_i$, 
$b_1$, $c_i$ and $d_i$ there exists a choice of the parameters and total 
quantities of the enzymes for the MM-MA system which gives rise to them. 
This choice can be made so as to depend smoothly on $(a_i,b_1,c_i,d_i)$.

In this notation, the Michaelis-Menten system for the truncated cascade
is the following:
\begin{eqnarray}
&&\frac{d}{dt}(\overline{KKK})=
\frac{c_1(KKK_{\rm tot}-\overline{KKK})}{1+d_1(KKK_{\rm tot}-\overline{KKK})}\nonumber\\
&&-\frac{c_2\overline{KKK}}
{1+d_2\overline{KKK}+b_1(KK_{\rm tot}-[KKPP]},\label{mm1}\\
&&\frac{d}{dt}[KK]=-\frac{a_1\overline{KKK}[KK]}
{1+b_1(KK_{\rm tot}-[KKPP])}\nonumber\\
&&+\frac{a_2(KK_{\rm tot}-[KK]-[KKPP])}
{1+b_1(KK_{\rm tot}-[KK])},\label{mm2}\\
&&\frac{d}{dt}[KKPP]=+\frac{a_3\overline{KKK}(KK_{\rm tot}-[KK]-[KKPP])}
{1+b_1(KK_{\rm tot}-[KKPP]}\nonumber\\
&&-\frac{a_4[KKPP]}
{1+b_1(KK_{\rm tot}-[KKPP])}.\label{mm3}
\end{eqnarray}
\section{Analysis of the truncated system}

The aim of this section is to prove that there are values of the parameters
and the total amounts of substrates and enzymes in the truncated Huang-Ferrell 
system for which there exist periodic solutions. The first step is to 
determine explicit stationary solutions of the system (\ref{mm1})-(\ref{mm3}) 
for certain values of the total amounts. The existence of periodic 
solutions of the MM system is obtained by showing that a Hopf bifurcation 
occurs at these stationary solutions. It is then shown that there is also a 
Hopf bifurcation, and hence periodic solutions, for the full truncated 
Huang-Ferrell system. Suppose that all parameters $a_i$, $b_1$, $c_2$ and $d_i$ 
have been fixed. Consider stationary solutions of (\ref{mm1})-(\ref{mm3}) 
which satisfy $[KK]=[KKPP]$. Then 
$a_1\overline{KKK}[KK]=a_2[KKP]$ and $a_3\overline{KKK}[KKP]=a_4[KKPP]$
as a consequence of (\ref{mm2}) and (\ref{mm3}). 
In particular the product of these two equations gives
$\overline{KKK}^2=\frac{a_2a_4}{a_1a_3}$ and so $\overline{KKK}$
is determined. Next it is possible to determine $[KK]$ and $[KKP]$ as
functions of $KK_{\rm tot}$ by means of the formulae
\begin{equation}
[KK]=\frac{KK_{\rm tot}}{2+q_1},\ \ \ [KKP]=\frac{q_1KK_{\rm tot}}{2+q_1},
\end{equation} 
where $q_1=\sqrt{\frac{a_1a_4}{a_2a_3}}$. 
The first equation (\ref{mm1}) has not been used yet and  produces a constraint on the parameters. This is the price to pay for simplifying some computations by having an equilibrium whose two last coordinates are identical. 
Set 
$Q=\frac{c_2\overline{KKK}}{1+d_2\overline{KKK}
+b_1(KK_{\rm tot}-[KK])}$. It follows from the equation $d\overline{KKK}/dt=0$
that 
$\frac{c_1(KKK_{\rm tot}-\overline{KKK})}
{1+d_1(KKK_{\rm tot}-\overline{KKK})}=Q$. This is equivalent to
$KKK_{\rm tot}=\overline{KKK}+\frac{Q}{c_1-d_1Q}$ provided the 
denominator in the last term is positive. Choose $c_1$ so that
$\frac{c_2d_1}{c_1d_2}<1$. Then the denominator is positive and this 
expression can be used to define $KKK_{\rm tot}$. Once this has been done all 
the conditions for stationary solutions are satisfied. Let us summarize this information:
\begin{equation} \label{equill}
\begin{array}{ll}
\mbox{coordinates of the equilibrium}& \overline{KKK}=\sqrt{\frac{a_2 a_4}{a_1 a_3}} \nonumber\\
& [KK]=[KKPP]=\frac{KK_{\rm tot}}{2+\sqrt{\frac{a_1 a_4}{a_2 a_3}}}\\
\mbox{constraints on the parameters}& \frac{c_2 d_1}{c_1 d_2}<1 \\
&KKK_{\rm tot}=\sqrt{\frac{a_2 a_4}{a_1 a_3}}+\frac{Q}{c_1 - d_1 Q},\\
& \mbox{where } Q=\frac{c_2\overline{KKK}}{1+d_2\overline{KKK}
+b_1(KK_{\rm tot}-[KK])} 
\end{array}
\end{equation}

Next the linearization of (\ref{mm1})-(\ref{mm3}) will be considered. It 
turns out that the signs of all entries in the derivative of the right hand
side of the system are independent of the values of the concentrations and can 
be determined. One is zero, one is positive and all the rest are negative. The 
characteristic polynomial of the linearization is the determinant of a matrix 
of the form
\begin{equation}\label{lin}
\left[
{\begin{array}{ccc}
\lambda+a & 0 & c \\ 
d & \lambda+e & f \\ 
-g & h & \lambda+i  
\end{array}}
\right]
\end{equation}
where all parameters are positive. Note that $-e$, $-f$, $-h$ and $-i$ only 
differ from the elements of the matrix (46) of \cite{hell14} by the 
multiplicative factor $N^{-2}$ and the fact that $a_1$ and $a_3$ are replaced 
by $a_1\overline{KKK}$ and $a_3\overline{KKK}$, respectively. As in that case 
the parameters $a_1\overline{KKK}$ and $a_4$ can be eliminated. 
Let $M_2$ be the submatrix of the matrix (\ref{lin}) with $\lambda=0$ 
consisting of the second and third rows and columns. 
The determinant of the linearization is given by 
\begin{equation}
-\left( a\vert M_2 \vert + c(dh+eg)\right).
\end{equation}
Since we are seeking parameters for which a Hopf bifurcation producing stable periodic orbits takes place, the three eigenvalues of the linearization at the equilibrium (\ref{equill}) should be two imaginary eigenvalues $\pm i\omega_0$, $\omega_0\neq 0$, and a third real negative eigenvalue. In particular the determinant of the linearization and its trace should be both strictly negative. The sign of $\vert M_2 \vert$ is unclear, but the linearization has a negative determinant if $\vert M_2 \vert$ is small enough.
Define $u=b_1KK_{\rm tot}$ and $v=b_1[KK]$. Then the determinant of $M_2$ is
given, up to a positive multiplicative constant, by  
\begin{equation}
-(u-3v+1)(u-2v)u(u-v+1).
\end{equation}
In fact all factors except the first are positive for all allowed values of the
variables. Thus the determinant is negative iff $u-3v+1>0$. This is equivalent
to the condition $ei-fh<0$. In terms of the parameters of the reduced system
$u-3v+1=\left(\frac{q_1-1}{q_1+2}b_1KK_{\rm tot}+1)\right)$. It follows that 
provided $q_1<1$ the value of $KK_{\rm tot}$ can be varied so that the 
determinant of $M_2$ passes from being positive to being negative. 

Now consider the determinant of the matrix in the linearized 
system. It is equal to $-a(ei-fh)-c(dh+eg)$. The second term is automatically 
negative and so if $ei-fh>0$ the determinant of the full matrix is negative. 
On the other hand if $ei-fh<0$ the sign of the determinant is not clear. The 
trace is negative for all parameter values. In a deformation as above the 
determinant of the three-dimensional matrix is negative when the determinant
of $M_2$ becomes zero. It stays negative at least in a small neighbourhood of 
that point. Thus there are parameters for which both the determinant of $M_2$
and the determinant of the three-dimensional matrix are negative. The 
determinant of (\ref{lin}) is
\begin{equation}
\lambda^3+(a+e+i)\lambda^2+(ae+ai+cg+ei-fh)\lambda+(aei-afh+cdh+cge).
\end{equation}
Call it $p_3(\lambda)$.
The eigenvalue conditions for a Hopf bifurcation are that two eigenvalues are 
purely imaginary and non-zero and that the third is non-zero. If this is true 
the trace is given by the real eigenvalue and thus determines its sign. In the 
example here we see that at a Hopf bifurcation the real eigenvalue must be 
negative. The determinant is then also negative. Introduce the notation
$p_3(\lambda)=\lambda^3+A_2\lambda^2+A_1\lambda+A_0$. In a deformation as above
all $A_i$ are positive near the point where $ei-fh=0$.  The Routh-Hurwitz 
theorem \cite{gantmacher59} implies that all roots of this polynomial have 
negative real parts if and only if $A_i>0$ for all $i$ and $H_3=A_2A_1-A_0>0$. 
The Routh-Hurwitz coefficient $H_3$ is given by  
\begin{equation}
H_3=(a+e+i)(ae+ai+cg+ei-fh)-aei+afh-cdh-cge
\end{equation}
which simplifies to 
\begin{equation}\label{hopf}
H_3=a(ae+ai+cg)+(e+i)(ae+ai+ei)+cgi-[cd+(e+i)f]h.
\end{equation}
Let $S_3$ be the region in the space of three by three matrices defined by the
condition that all eigenvalues have negative real parts. At any point on the 
boundary of $S_3$ where the determinant is non-zero the eigenvalue 
configuration characteristic of the Hopf bifurcation must occur. In other 
words, two eigenvalues are imaginary and non-zero and the other is negative.

In the Michaelis-Menten system for the truncated MAPK cascade we have already 
exhibited a class of stationary solutions and shown that there is a choice of
the parameters for which the matrix defined by the linearization of this 
system has negative determinant and the determinant of $M_2$ is negative. Call 
this point in parameter space $Z$.
We recall that at $Z$, the constraints that have to be fulfilled are the following:
\begin{equation}\label{constraintss}
\begin{array}{lll}
\frac{c_2 d_1}{c_1 d_2}&<&1\\
KKK_{\rm tot}&=& \sqrt{\frac{a_2 a_4}{a_1 a_3}}\\
q_1&=&\sqrt{\frac{a_1 a_4}{a_2 a_3}}<1\\
KK_{\rm tot}&=& \frac{q_1+2}{b_1(1-q_1)}+\delta,
\end{array}
\end{equation}
where $\delta>0$ is small.
 At $Z$ the linearization has negative 
trace. The coefficients will now be rescaled in a particular way. Let 
$\bar a_i$ be the values of the parameters $a_i$ at the starting point and 
define $a_i(L)=\bar a_i L$. The other parameters are kept fixed. Under this 
rescaling a stationary solution remains a stationary solution. The determinant 
of the linearization is rescaled by $L^2$ and so, in particular, has a fixed 
sign. The constants $d$-$i$ in the matrix of the linearization are scaled by a 
factor $L$ while the coefficients $a$ and $c$ are left unchanged. After 
rescaling the condition that the quantity $H_3$ is negative becomes
\begin{equation}
L^3(fh-ei)(e+i)>La(ae+ai+cg)+L^2(e+i)(ae+ai)+L^2cgi-L^2cdh.
\end{equation}
If we start the rescaling from $Z$ then for $L$ sufficiently large this 
condition will be satisfied. On the other hand, for $L$ sufficiently
small and positive the opposite inequality will be satisfied. In other words, 
in this family the quantity $H_3$ starts positive for $L$ small and positive 
and becomes negative at some large value of $L$. It must pass from being 
positive to being zero at some point $L_0$. In fact it must have a minimum for 
some negative value of $L$ and a maximum for some positive value of $L$ since 
it vanishes at the origin and its derivative there is positive. In fact the value $L_0$ is the unique positive root of a quadratic polynomial.  At the unique 
positive value $L_0$  of $L$ for which it is zero this function has a negative 
derivative.  It follows from the facts that $A_2>0$ and $A_0>0$ that $A_1$ 
cannot approach zero before $H_3$ reaches zero. Thus the family cannot leave 
$S_3$ before $H_3$ reaches zero. This family passes through a point where the 
eigenvalues satisfy the conditions for a Hopf bifurcation. When a 
three-dimensional matrix has one negative real eigenvalue $\mu_1$ and another 
eigenvalue $\mu_2$ which is not real the quantity $H_3$ is equal to
\begin{equation}
-{\rm Re}\mu_2[({\rm Re}\mu_2)^2+2({\rm Im}\mu_2)^2+(\mu_1+{\rm Re}\mu_2)^2].
\end{equation} 
From this we can see that if a one-parameter family of matrices is such that 
the quantity $H_3$ passes through zero at some parameter value with 
non-zero velocity then the real part of the complex eigenvalue passes through
zero at that point with non-zero derivative. 

The situation which has just been described is as follows. We have a family of 
coefficients in the Michaelis-Menten system depending on the parameter $L$. 
There is a corresponding family of stationary solutions. We consider the 
linearization of the system at that point. We can suppose that for all values
of $L$ considered it has one negative real eigenvalue $\mu_1$ and one 
eigenvalue $\mu_2$ which is never real. Moreover at $L=L_0$ the eigenvalue
$\mu_2$ is imaginary and its real part moves through the imaginary axis with
non-zero velocity as $L$ is varied through $L_0$. In other words, $\mu_2 (L)$
defines a curve in the complex plane which passes through the imaginary 
axis transversely for $L=L_0$. In particular the linearization is always 
invertible.
It was shown in \cite{hell14} that the Michaelis-Menten system
(\ref{mm1})-(\ref{mm3}) can be embedded into the MM-MA system 
(\ref{tev1})-(\ref{tev9}) in such a way as to allow GSPT to be applied. In the
terminology of \cite{hell14} the transverse eigenvalues all have negative real
parts. More information on these matters can be found in the appendix. Above
we introduced a parameter $L$ into the MM system by letting the coefficients
depend on $L$ in a certain way. This can be extended to the MM-MA system by
making a smooth choice of corresponding coefficients for that system.
After 
doing this the system on the slow manifold depends on the two parameters 
$\epsilon$ and $L$. At the point where the Hopf bifurcation occurs for 
$\epsilon=0$ and $L=L_0$ the derivative of the right hand side of the 
system with respect to the unknowns is invertible. Hence, by the implicit 
function theorem there is a unique nearby stationary solution depending 
smoothly on $L$ and $\epsilon$. Thus the family of stationary solutions for
$\epsilon=0$ can be continued smoothly to a curve of stationary solutions in 
the slow manifold for small positive values of $\epsilon$. For each $\epsilon$ 
there are corresponding eigenvalues 
$\tilde\mu_1 (L,\epsilon)=\mu_{1,\epsilon} (L)$ and 
$\tilde\mu_2 (L,\epsilon)=\mu_{2,\epsilon} (L)$ of the linearization at the
stationary solution. For $\epsilon$ small the curve $\mu_{2,\epsilon}$ intersects 
the imaginary axis transversely. Now the transverse eigenvalues are negative 
and it follows that the family of solutions of the MM-MA system 
undergoes a Hopf bifurcation. Using centre manifold theory the dynamics can 
be reduced to a two-dimensional situation for both the MM-MA system for the
truncated Huang-Ferrell model and the corresponding MM system. Applying the 
theorem of Hopf in the form given in \cite{hale69}, Chapter VIII, Theorem 1.3 
shows that in both cases there exists a one-parameter family of periodic 
solutions. The following theorem has been proved.
\begin{theorem}\label{theorem1}
There exist positive parameter values $a_i$, $b_1$, $c_i$ and 
$d_i$ such that the MM system for the truncated Huang-Ferrell model with 
these parameter values has a positive periodic solution. There exist positive
parameter values $\tilde a_i$, $k_i$ and $\tilde d_i$ and positive values of 
the total amounts $E_{1,\rm tot}$, $E_{2,\rm tot}$, $(KKP'ase)_{\rm tot}$, 
$KKK_{\rm tot}$ and $KK_{\rm tot}$ such that the MM-MA system for the truncated 
Huang-Ferrell model with these parameter values and these values of the 
total amounts has a positive periodic solution.
\end{theorem}

\section{Bistability}\label{bistab}

In \cite{hell14} it was proved that there are parameter values for which the 
dual futile cycle has more than one stable stationary solution. It will now 
be shown that this property is inherited by the truncated MAPK cascade. For 
this the coefficients in the MM system for the truncated MAPK cascade will be
rescaled. Define $c_1=\epsilon^{-1}\hat c_1$, $c_2=\epsilon^{-2}\hat c_2$
and $d_2=\epsilon^{-1}\hat d_2$. Then the equations for the quantities with hat 
are in the standard form for singular perturbation theory. For fixed values of 
$\overline{KKK}$ the equations for $[KK]$ and $[KKPP]$ are those for the dual 
futile cycle and we know that there are parameters for which there exist two 
stable hyperbolic stationary solutions. The evolution equation for 
$\overline{KKK}$ is
\begin{eqnarray}
&&\epsilon\frac{d}{dt}(\overline{KKK})
=\frac{\hat c_1(KKK_{\rm tot}-\overline{KKK})}
{1+d_1(KKK_{\rm tot}-\overline{KKK})}\nonumber\\
&&-\frac{\hat c_2\overline{KKK}}
{\epsilon+\hat d_2\overline{KKK}+\epsilon b([KK]+[KKP])}.
\end{eqnarray} 
For $\epsilon=0$ this reduces to 
\begin{equation}
d_1\left(\frac{\hat c_1\hat d_2}{\hat c_2 d_1}-1\right)
(KKK_{\rm tot}-\overline{KKK})=1.
\end{equation} 
Interestingly this last equation does not depend on $[KK]$ or $[KKP]$. 
Provided the quotient of reaction constants occurring in this equation, which 
is equal to $\frac{c_1d_2}{c_2d_1}$, is greater than one then a unique stationary
solution is determined by solving this equation for $KKK_{\rm tot}$ in terms of 
$\overline{KKK}$. It is an asymptotically stable and hyperbolic solution of
the equation
\begin{equation}
\frac{d}{dt}(\overline{KKK})=\frac{\hat c_1(KKK_{\rm tot}-\overline{KKK})}
{1+d_1(KKK_{\rm tot}-\overline{KKK})}
-\frac{\hat c_2}{\hat d_2}.
\end{equation}
Thus the 
transverse eigenvalue in the sense of GSPT is negative and that theory can be
applied. It follows that the MM system for the truncated MAPK cascade 
exhibits bistability. This implies a corresponding statement for the MM-MA 
system. It can be shown in a similar way that both the MM and MM-MA systems 
contain saddle points for these values of the parameters. 

It can also be shown that there are parameters for which there is a stationary 
value where the  eigenvalues are real with the sign pattern $(-,-,0)$. There
$ei-fh<0$. Scaling with the parameter $L$ starting at this point gives a 
family of coefficients for which the determinant of the linearization is 
always zero. Along this family the quantity $H_3$ goes from positive to
negative. When it crosses zero $A_2=0$ and there is a second zero eigenvalue. 
If the kernel of the matrix at that point were two-dimensional it would have 
to intersect the subspace spanned by the first two coordinates in a 
one-dimensional subspace. It would follow that this subspace was spanned by the 
vector with components $(0,1,0)$. This gives a contradiction. Thus the Jordan
form of the matrix must be non-diagonalizable and the algebraic conditions 
on the linearization for a Bogdanov-Takens bifurcation are satisfied at that 
point. In other words, the condition BT.0 of \cite{kuznetsov10} holds. Since,
on the other hand, no information has been obtained on the genericity 
conditions BT.1-BT.3 this does not by itself give useful information on the 
dynamical properties of the solutions.

\section{The full Huang-Ferrell system}

The aim is now to reduce the full Huang-Ferrell system in a way similar to 
that done above for the truncated system. The first step is to introduce the
quantity
\begin{equation}
\overline {KK}=[KKPP]+[K\cdot KKPP]+[KP\cdot KKPP].
\end{equation}
Consider the system of evolution equations for $\overline{KKK}$, $[KK]$, 
$\overline{KK}$, $[K]$, $[KPP]$ and the substrate-enzyme complexes and rescale
the unknowns. In this case the quantities not containing
the $E_i$, $KKK$, $KK$, $KKP'ase$ or $KP'ase$ are not rescaled. These are the
first three quantities in the fourth line of (\ref{substances}). Quantities 
which contain $KK$ or $KP'ase$ but not $E_i$, $KKK$ or $KKP'ase$ are rescaled 
by $\epsilon^{-1}$. These are the first three quantities in the second line,
the last quantity in the fourth line and all quantities in the fifth line of 
(\ref{substances}). Quantities which contain $KKK$ or $KKP'ase$ but not $E_i$ 
are rescaled by $\epsilon^{-2}$. These are the first two quantities in the first
line, the last quantity in the second line and all quantities in the third 
line of (\ref{substances}). Quantities which contain the $E_i$ are rescaled by 
$\epsilon^{-3}$. These are the last four quantities in the first line of 
(\ref{substances}). The reaction constants $\tilde a_1$ and $\tilde a_2$ are 
multiplied by $\epsilon^2$ to get new quantities while $\tilde a_3$, 
$\tilde a_4$, $\tilde a_5$ and $\tilde a_6$ are multiplied by $\epsilon$. A 
new time coordinate is introduced as $\epsilon$ times the old one. These 
scalings have been chosen so that the new equations for the free substrates 
are independent of $\epsilon$ and the new equations for the substrate-enzyme 
complexes have a factor $\epsilon$ in front of the time derivatives, just as 
in the case of the truncated system. The resulting equations extend 
(\ref{tev1})-(\ref{tev9}). The first two rescaled equations are unchanged and 
the third only differs from the corresponding equation for the truncated 
system in that $[KKPP]$ is replaced by $\overline{KK}$ on the left hand side. 
The list of expressions for the concentrations of the complexes can be 
extended as follows
\begin{eqnarray}
&&[K\cdot KKPP]=\frac{\tilde a_7[K][KKPP]}{\tilde d_7+k_7},\\
&&[KP\cdot KP'ase]=\frac{\tilde a_8[KP][KP'ase]}{\tilde d_8+k_8},\\
&&[KP\cdot KKPP]=\frac{\tilde a_9[KP][KKPP]}{\tilde d_9+k_9},\\
&&[KPP\cdot KP'ase]=\frac{\tilde a_{10}[KPP][KP'ase]}{\tilde d_{10}+k_{10}}.
\end{eqnarray}
There are the following additional equations for the total amounts of 
enzymes
\begin{eqnarray}
&&\overline{KK}=[KKPP](1+K_{m,7}^{-1}[K]+K_{m,9}^{-1}[KP]),\\
&&(KP'ase)_{\rm tot}=[KP'ase](1+K_{m,8}^{-1}[KP]+K_{m,10}^{-1}[KPP]).
\end{eqnarray}
The equations (\ref{inter1})-(\ref{inter3}) can be taken over to the full
model except that in (\ref{inter3}) the quantity $[KKPP]$ should be replaced
by $\overline{KK}$ on the left hand side and that in order to obtain a closed
system $[KKPP]$ must be substituted for in terms of $\overline{KK}$ on the 
right hand side. The following equations also hold:
\begin{eqnarray}
&&\frac{d}{dt}[K]=-\frac{k_7}{K_{m,7}}[K][KKPP]
+\frac{k_8}{K_{m,8}}[KP][KP'ase],\label{inter4}\\
&&\frac{d}{dt}[KPP]=-\frac{k_{10}}{K_{m,10}}[KPP][K P'ase]
+\frac{k_9}{K_{m,9}}[KP][KKPP].\label{inter5}
\end{eqnarray} 
The evolution equations for the MM system are
\begin{eqnarray}
&&\frac{d}{dt}(\overline{KKK})=
\frac{k_1K_{m,1}^{-1}E_{1,{\rm tot}}[KKK]}{1+K_{m,1}^{-1}[KKK]}\nonumber\\
&&-\frac{k_2K_{m,2}^{-1}E_{2,{\rm tot}}\overline{KKK}}
{1+K_{m,2}^{-1}\overline{KKK}+K_{m,3}^{-1}[KK]+K_{m,5}^{-1}[KKP]},\label{mm4}\\
&&\frac{d}{dt}[KK]=-\frac{k_3K_{m,3}^{-1}\overline{KKK}[KK]}
{1+K_{m,3}^{-1}[KK]+K_{m,5}^{-1}[KKP]}\nonumber\\
&&+\frac{k_4K_{m,4}^{-1}(KK P'ase)_{\rm tot}[KKP]}
{1+K_{m,4}^{-1}[KKP]+K_{m,6}^{-1}[KKPP]},\label{mm5}\\
&&\frac{d}{dt}(\overline{KK})=+\frac{k_5K_{m,5}^{-1}\overline{KKK}[KKP]}
{1+K_{m,3}^{-1}[KK]+K_{m,5}^{-1}[KKP]}\nonumber\\
&&-\frac{k_6K_{m,6}^{-1}(KK P'ase)_{\rm tot}[KKPP]}
{1+K_{m,4}^{-1}[KKP]+K_{m,6}^{-1}[KKPP]},\label{mm6}\\
&&\frac{d}{dt}[K]=-\frac{k_7K_{m,7}^{-1}\overline{KK}[K]}
{1+K_{m,7}^{-1}[K]+K_{m,9}^{-1}[KP]}\nonumber\\
&&+\frac{k_8K_{m,8}^{-1}(K P'ase)_{\rm tot}[KP]}
{1+K_{m,8}^{-1}[KP]+K_{m,10}^{-1}[KPP]},\label{mm7}\\
&&\frac{d}{dt}[KPP]=+\frac{k_9K_{m,9}^{-1}\overline{KK}[KP]}
{1+K_{m,7}^{-1}[K]+K_{m,9}^{-1}[KP]}\nonumber\\
&&-\frac{k_{10}K_{m,10}^{-1}(K P'ase)_{\rm tot}[KPP]}
{1+K_{m,8}^{-1}[KP]+K_{m,10}^{-1}[KPP]}.\label{mm8}
\end{eqnarray}    
Assume that the coefficients $K_{m,i}^{-1}$ are equal to a common quantity $b_2$ 
for $7\le i\le 10$. Extend the definition of $a_i$ by defining it to be
$k_{i+2}K_{m,i+2}^{-1}$ for $i=5,7$ and $k_{i+2}K_{m,i+2}^{-1}(KP'ase)_{\rm tot}$ for
$i=6,8$. As in the case of the truncated system there exists a smooth choice 
of the parameters and total quantities of the enzymes for the MM-MA system 
which give rise to any choice of the parameters $a_i$, $b_i$, $c_i$ and $d_i$. 

We now have a system describing the full cascade which is in the standard 
form of GSPT. To profit from this it is necessary to examine the transverse
eigenvalues. These are the eigenvalues of the matrix which is the derivative of 
right hand side of the evolution equations for the enzyme-substrate complexes
for fixed values of the substrate concentrations. There are ten complexes and 
this is a ten by ten matrix with components $L_{ij}$. Let the complexes be 
numbered in the order they are listed in (\ref{substances}). Each complex 
which does not share an enzyme contributes a diagonal element to the matrix.
These are the components $L_{11}$ and $L_{22}$ and are negative. Each pair of 
complexes which share an enzyme contributes a two by two submatrix on the 
diagonal. They are the $L_{ij}$ with $2k-1\le i,j\le 2k$ for $2\le k\le 5$.
The eigenvalues of each of these submatrices on the diagonal have negative real 
parts. The calculation is essentially the same as that done for the case of
the truncated system in \cite{hell14}. It remains to examine the effect of the 
non-zero elements of the matrix which do not belong to any of these blocks.
These are $L_{23}$, $L_{24}$, $L_{67}$ and $L_{68}$. The elements $L_{11}$ and 
$L_{22}$ are alone in their columns. Thus the calculation of the eigenvalues
reduces to that of the submatrix obtained by discarding the first and second 
rows and columns. Then the submatrices for $k=1$ and $k=4$ occur
as direct sums with other matrices. Thus determining the eigenvalues reduces
to doing so for the submatrix defined by $5\le i,j\le 8$. This submatrix
is block upper triangular and so its eigenvalues are the eigenvalues of the 
submatrices for $k=2$ and $k=3$. Combining these facts shows that all 
transverse eigenvalues have negative real parts. 

Consider a stationary solution of the system (\ref{mm4})-(\ref{mm8}) which 
satisfies $[KK]=[KKPP]$ and $[K]=[KPP]$. Explicit stationary solutions
can be found in a way similar to what was done for the truncated system. 
It follows from equations (\ref{mm7}) and (\ref{mm8}) that
$a_5\overline{KK}[K]=a_6[KP]$ and $a_7\overline{KK}[KP]=a_8[KPP]$. Hence
$\overline{KK}^2=\frac{a_6a_8}{a_5a_7}$. The quantities $[K]$ and $[KP]$ are
determined by 
\begin{equation}
[K]=\frac{K_{\rm tot}}{2+q_2},\ \ \ [KP]=\frac{q_2K_{\rm tot}}{2+q_2},
\end{equation} 
where $q_2=\sqrt{\frac{a_5a_8}{a_6a_7}}$. The expression obtained for 
$\overline{KKK}$ in the case of the truncated system remains valid for the 
full system while in those for $[KK]$ and $[KKP]$ are modified to
\begin{equation}
[KK]=\frac{KK_{\rm tot}}{2+\tilde q_1},\ \ \ 
[KKP]=\frac{q_1KK_{\rm tot}}{2+\tilde q_1},
\end{equation} 
where $\tilde q_1=q_1+b_2(K_{\rm tot}-[K])$. In addition we have the relation
\begin{equation}
KK_{\rm tot}=\frac{(2+\tilde q_1)\overline{KK}}{1+b_2(K_{\rm tot}-[K])},
\end{equation}
so that $KK_{\rm tot}$ is determined. Stationary solutions for the MM system
for the full cascade can be determined as follows. Fix the parameters $a_i$, 
$b_i$, $c_2$, $d_i$. Then if $c_1$ is chosen sufficiently large the 
concentrations for the stationary solution can be expressed in terms of 
$K_{\rm tot}$ and $KK_{\rm tot}$. Note that if the parameters $a_i$, $1\le i\le 8$
are varied in such a way that $r_k=a_{2k+1}/a_{2k}$ remains unchanged for 
$1\le k\le 4$ then $q_1$ and $q_2$ do not change and the stationary solution
is preserved.

Expressing the evolution equations in terms of the quantities $a_i$, $b_i$, 
$c_i$ and $d_i$ and using the conservation laws gives
\begin{eqnarray}
&&\frac{d}{dt}(\overline{KKK})=
\frac{c_1(KKK_{\rm tot}-\overline{KKK})}
{1+d_1(KKK_{\rm tot}-\overline{KKK})}\nonumber\\
&&-\frac{c_2\overline{KKK}}
{1+d_2\overline{KKK}+b_1(KK_{\rm tot}-\overline{KK})},\label{mapkmm1}\\
&&\frac{d}{dt}[KK]=-\frac{a_1\overline{KKK}[KK]}
{1+b_1(KK_{\rm tot}-\overline{KK})}\nonumber\\
&&+\frac{a_2(KK_{\rm tot}-[KK]-\overline{KK})}
{1+b_1\left(KK_{\rm tot}-[KK]-\frac{b_2(K_{\rm tot}-[KPP])}
{1+b_2(K_{\rm tot}-[KPP])}\overline{KK}\right)},\label{mapkmm2}\\
&&\frac{d}{dt}(\overline{KK})=
\frac{a_3\overline{KKK}(KK_{\rm tot}-[KK]-\overline{KK})}
{1+b_1(KK_{\rm tot}-[KK])}\nonumber\\
&&-\frac{a_4\overline{KK}}
{b_1\overline{KK}+(1+b_1(KK_{\rm tot}-[KK]-\overline{KK}))
(1+b_2(K_{\rm tot}-[KPP]))},\label{mapkmm3}\\
&&\frac{d}{dt}[K]=-\frac{a_5\overline{KK}[K]}
{1+b_2(K_{\rm tot}-[KPP])}\nonumber\\
&&+\frac{a_6(K_{\rm tot}-[K]-[KPP])}
{1+b_2(K_{\rm tot}-[K])},\label{mapkmm4}\\
&&\frac{d}{dt}[KPP]=+\frac{a_7\overline{KK}(K_{\rm tot}-[K]-[KPP])}
{1+b_2(K_{\rm tot}-[KPP])}\nonumber\\
&&-\frac{a_{8}[KPP]}
{1+b_2(K_{\rm tot}-[K])}.\label{mapkmm5}
\end{eqnarray}     
A relation will now be established between this system and the system for 
the truncated model by doing some rescaling. Replace $[K]$, $[KPP]$ and 
$K_{\rm tot}$ by $\epsilon [K]$, $\epsilon [KPP]$ and $\epsilon K_{\rm tot}$,
respectively. Replace $a_i$ by $\epsilon^{-1} a_i$ for $5\le i\le 8$. In the 
limit the first three equations are independent of $[K]$ and $[KPP]$ and are 
just the equations of the truncated system analysed in the last section. The 
limit is in the form appropriate for applying GSPT. The linearization of 
the system for $[K]$ and $[KPP]$ is independent of those two variables. Its
trace and determinant are $-(a_5+a_7)\overline{KK}-(a_6+a_8)<0$ and 
$a_5a_7\overline{KK}^2+(a_5+a_6)a_8>0$, respectively and so the transverse 
eigenvalues have negative real parts. A parameter $L$ can be introduced in the 
system for the full cascade in the same way as was done for the truncated 
cascade. It follows that the presence of a Hopf bifurcation in the MM system 
for the truncated model implies that of Hopf bifurcation in the MM system for 
the full cascade. A parameter $L$ can also be introduced in the MM-MA system 
for the full cascade, implying the existence of a Hopf bifurcation for that 
system, i.e. the original system of Huang and Ferrell. Thus the following 
theorem has been proved.
\begin{theorem}\label{theorem2}
There exist positive parameter values $a_i$, $b_i$, $c_i$ and 
$d_i$ such that the MM system for the full Huang-Ferrell model with 
these parameter values has a positive periodic solution. There exist positive
parameter values $\tilde a_i$, $k_i$ and $\tilde d_i$ and positive values of 
the total amounts $E_{1,\rm tot}$, $E_{2,\rm tot}$, $(KKP'ase)_{\rm tot}$, 
$(KKP'ase)_{\rm tot}$, $KKK_{\rm tot}$, $KK_{\rm tot}$ and $K_{\rm tot}$ such that the 
MM-MA system for the full Huang-Ferrell model with these parameter values and 
these values of the total amounts has a positive periodic solution.
\end{theorem}

%

\section{Further examples}

This section is concerned with some examples which are variations on those
coming from the MAPK cascade. The first is a system which is similar to the 
truncated Huang-Ferrell system except that the second layer of the cascade 
only has one phosphorylation site. 
In other words, we discuss now the following cascade:
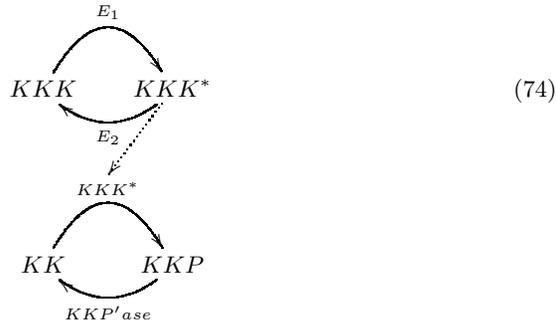
\begin{equation}\label{cascademonophos}
\xymatrix@C=0.5em{
 KKK\ar@/^2pc/[rr]^{E_1}&&KKK^* \ar@/^1pc/[ll]^{E_2}   \ar@{.>}[dl] &&&&\\
 &&&&&&\\
KK\ar@/^2pc/[rr]^{KKK^*}&&KKP \ar@/^1pc/[ll]^{KKP'ase}
}
\end{equation}
This minimal cascade and its generalization 
to several layers of simple phosphorylation loops have been considered in
\cite{ventura08} where it is remarked that damped oscillations may occur in
a system of this type. It will now be shown how the features observed for the 
MAPK cascade change in the case of a cascade of two simple phosphorylation 
loops. The variables are 
\begin{eqnarray}
&&[KKK], [KKK^*], [E_1], [E_2], [KKK\cdot E_1], [KKK^*\cdot E_2],
\nonumber\\  
&&[KK], [KKP], [KKP'ase], [KK\cdot KKK^*], [KKP\cdot KKP'ase]. 
\end{eqnarray}
We introduce the variable
$\overline{KKK}=[KKK^*]+[KK\cdot KKK^*]$ in analogy to what was done in the 
previous examples. The variables are rescaled as in the truncated 
Huang-Ferrell model. This means that $[KK]$ and $[KKP]$ are not rescaled, 
for $[KKK]$, $[KKK^*]$, $[KKP'ase]$, $[KK\cdot KKK^*]$ and 
$[KKP\cdot KKP'ase]$ the new quantity is $\epsilon^{-1}$ times the old one
while for $[E_1]$, $[E_2]$, $[KKK\cdot E_1]$ and $[KKK^*\cdot E_2]$ the new 
quantity is $\epsilon^{-2}$ times the old one. The reaction constants 
$\tilde a_1$ and $\tilde a_2$ are multiplied by $\epsilon$ to get new 
quantities. A new time coordinate is introduced as $\epsilon$ times the old
one. There results the following system
\begin{eqnarray}
&&\frac{d}{dt}(\overline{KKK})=-\tilde a_2[KKK^*][E_2]+d_2[KKK^*\cdot E_2]
+k_1[KKK\cdot E_1],\\
&&\frac{d}{dt}[KK]=-\tilde a_3[KK][KKK^*]+d_3[KK\cdot KKK^*]
+k_4[KKP\cdot KK P'ase],\\
&&\epsilon\frac{d [KKK\cdot E_1]}{dt}=\tilde a_1[KKK][E_1]
-(d_1+k_1)[KKK\cdot E_1],\\
&&\epsilon\frac{d [KKK^*\cdot E_2]}{dt}=\tilde a_2[KKK^*][E_2]-(d_2+k_2)
[KKK^*\cdot E_2],\\
&&\epsilon\frac{d [KK\cdot KKK^*]}{dt}=\tilde a_3[KK][KKK^*]-(d_3+k_3)
[KK\cdot KKK^*],\\
&&\epsilon\frac{d [KKP\cdot KK P'ase]}{dt}=\tilde a_4[KKP][KK P'ase]
-(d_4+k_4)[KKP\cdot KK P'ase].
\end{eqnarray} 
To get a closed system the following relations must be used.
\begin{eqnarray}
&&[KKK^*]=\overline{KKK}-[KK\cdot KKK^*],\\
&&[KKK]=KKK_{\rm tot}-\overline{KKK}+O(\epsilon),\\
&&[KKP]=KK_{\rm tot}-[KK]+O(\epsilon),\\
&&[E_1]=E_{1,{\rm tot}}-[KKK\cdot E_1],\\
&&[E_2]=E_{2,{\rm tot}}-[KKK^*\cdot E_2],\\
&&[KK P'ase]=(KK P'ase)_{\rm tot}-[KKP\cdot KK P'ase].
\end{eqnarray}
For $\epsilon=0$ the following MM system is obtained
\begin{eqnarray}
&&\frac{d}{dt}(\overline{KKK})=
\frac{k_1K_{m,1}^{-1}E_{1,{\rm tot}}[KKK]}{1+K_{m,1}^{-1}[KKK]}\nonumber\\
&&-\frac{k_2K_{m,2}^{-1}E_{2,{\rm tot}}\overline{KKK}}
{1+K_{m,2}^{-1}\overline{KKK}+K_{m,3}^{-1}[KK]},\label{mmm1}\\
&&\frac{d}{dt}[KK]=-\frac{k_3K_{m,3}^{-1}\overline{KKK}[KK]}
{1+K_{m,3}^{-1}[KK]}\nonumber\\
&&+\frac{k_4K_{m,4}^{-1}(KK P'ase)_{\rm tot}[KKP]}
{1+K_{m,4}^{-1}[KKP]}.\label{mmm2}
\end{eqnarray}
The system (\ref{mmm1})-(\ref{mmm2}) has the property that the trace of the 
derivative of the right hand side is always negative. Thus by the Dulac 
criterion this system admits no periodic solutions. It was shown in 
\cite{feliu12} that in this case the MM-MA system has a unique stationary
solution for given values of the parameters and this means that the 
same is true for the MM system. On the boundary of the region of positive
concentrations the vector field points inwards and all solutions are bounded 
due to the conservation laws. Putting all these facts together, it follows 
from Poincar\'e-Bendixson theory that the stationary solution of the MM 
system is globally asymptotically stable. It is not clear that stability
might not be lost for general values of the parameters in the MM-MA system. 

The next example is an in vitro model, introduced by Prabakaran, Gunawardena 
and Sontag \cite{prabakaran14}, of the MAPK cascade consisting of the proteins 
Raf, MEK and ERK. The model system is simplified compared to the original 
biological system in two ways. The protein Raf is constitutively active. 
This corresponds to taking a fixed value of $\overline{KKK}$ in the model. ERK 
is mutated so that it can only be phosphorylated once, on tyrosine and not on 
threonine. The role of the phosphatase $KKP'ase$ is played by PP2A (protein 
phosphatase 2A) and that of $KP'ase$ by PTP (protein tyrosine phosphatase). 
Normally PP2A can remove a phosphate group from the threonine in ERK, thus 
causing a mixing of the layers but the mutation ensures that a phosphate of 
this kind is not present and PP2A cannot remove the phosphate from tyrosine.
This leads to a cascade where the first layer allows two phosphorylation
steps but the second only allows one. 
This cascade is represented in the following diagram:
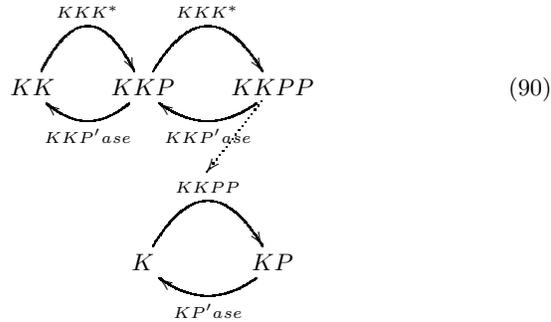
\begin{equation}\label{cascadepgs}
\xymatrix@C=0.5em{
 KK\ar@/^2pc/[rr]^{KKK^*}&&KKP \ar@/^1pc/[ll]^{KKP'ase} \ar@/^2pc/[rr]^{KKK^*}  && KKPP\ar@{.>}[dl]  \ar@/^1pc/[ll]^{KKP'ase}\\
 &&&&\\
&&K\ar@/^2pc/[rr]^{KKPP}&&KP \ar@/^1pc/[ll]^{KP'ase}
}
\end{equation}
Suppose that the phosphorylation and
dephosphorylation in the first layer are distributive and sequential. In other 
words, only one phosphate group is added or removed in each reaction, the 
groups are added in a specified order and removed in the reverse order. Given 
this data it is possible to set up an MM-MA model as done in other cases above 
but this is not the model used in \cite{prabakaran14}. There the 
phosphorylation of MEK is modelled using
mass action (MA) kinetics with both phosphates being added in one step and 
the concentration of Raf not included as a variable. On the other hand the
action of MEK as an enzyme in phosphorylating ERK is modelled in detail. 
This gives a kind of hybrid MA/MM-MA model, which we call the PGS model. It 
is proved in \cite{prabakaran14} that for the PGS model all solutions converge 
to a stationary solution at late times. 

Now the MM-MA model for the in vitro system of \cite{prabakaran14} will be 
examined, together with its MM reduction. In the Huang-Ferrell system discard 
the first four equations and take $\overline{KKK}$ to be a constant in the 
remaining equations. Then set $[KP\cdot KKPP]$, $[KPP]$ and 
$[KPP\cdot KP'ase]$ to zero together with the reaction constant $\tilde a_9$. 
Discard the last three equations. In the corresponding MM system this means 
taking $\overline{KKK}$ to be constant, setting $[KPP]$ and $a_7$ to zero and 
discarding the first and last equations. In this case the quantity 
$\overline{KK}$ is defined to be $[KKPP]+[K\cdot KKPP]$. It is no longer 
possible to keep the coefficients $K^{-1}_{m,i}$ with $7\le i\le 10$ equal to 
the same positive constant $b_2$. This can be required for $i\ne 9$ but 
$K^{-1}_{m,9}$ must be replaced by zero. This has the effect that the expression 
$b_2(K_{\rm tot}-[KPP])$ is replaced in equations (\ref{mapkmm2}), 
(\ref{mapkmm3}) and (\ref{mapkmm4}) by $b_2[K]$. The system of equations 
obtained is
\begin{eqnarray}
&&\frac{d}{dt}[KK]=-\frac{a_1\overline{KKK}[KK]}
{1+b_1(KK_{\rm tot}-\overline{KK})}\nonumber\\
&&+\frac{a_2(KK_{\rm tot}-[KK]-\overline{KK})}
{1+b_1\left(KK_{\rm tot}-[KK]-\frac{b_2[K]}
{1+b_2[K]}\overline{KK}\right)},\label{pgsmm1}\\
&&\frac{d}{dt}(\overline{KK})=
\frac{a_3\overline{KKK}(KK_{\rm tot}-[KK]-\overline{KK})}
{1+b_1(KK_{\rm tot}-[KK])}\nonumber\\
&&-\frac{a_4\overline{KK}}
{b_1\overline{KK}+(1+b_1(KK_{\rm tot}-[KK]-\overline{KK}))
(1+b_2[K])},\label{pgsmm2}\\
&&\frac{d}{dt}[K]=-\frac{a_5\overline{KK}[K]}
{1+b_2[K]}\nonumber\\
&&+\frac{a_6(K_{\rm tot}-[K])}
{1+b_2(K_{\rm tot}-[K])}.\label{pgsmm3}
\end{eqnarray} 

Next, in analogy with what has been done for other models above, stationary 
solutions will be considered which satisfy the restriction
$[KK]=[KKPP]$. These satisfy $a_1[KK]\overline{KKK}=a_2[KKP]$ and 
$a_3[KKP]\overline{KKK}=a_4[KK]$. Hence $a_1a_3\overline{KKK}^2=a_2a_4$ and 
$\overline{KKK}=\sqrt{\frac{a_2a_4}{a_1a_3}}$. Substituting this back in gives
$[KKP]=a_1a_2^{-1}\overline{KKK}[KK]=q_1[KK]$ where 
$q_1=\sqrt{\frac{a_1a_4}{a_2a_3}}$. Hence 
\begin{equation}
KK_{\rm tot}=[KK]+[KKP]+\overline{KK}=(2+q_1+b_2[K])[KK].
\end{equation}
This allows $[KK]$ and $[KKP]$ to be expressed in terms of $KK_{\rm tot}$, $q_1$,
$b_2$ and $[K]$. These relations are equivalent to the first and second 
equations for stationary solutions. The remaining equation can be written as
\begin{equation}
\overline{KK}=\frac{a_6(K_{\rm tot}-[K])(1+b_2[K])}{a_5[K](1+b_2(K_{\rm tot}-[K]))}.
\end{equation}
One way of determinining a set of stationary solutions is as follows. First 
choose $[K]$, $K_{\rm tot}>[K]$, $b_2$, and the $a_i$. Then use the last
equation to determine $\overline{KK}$. Next use
\begin{equation}
KK_{\rm tot}=\frac{(2+q_1+b_2[K])\overline{KK}}{1+b_2[K]}.
\end{equation}
Then $[KK]$, $[KKP]$ and $[KKPP]$ are determined in such a way that all 
the equations for stationary solutions are satisfied.
To summurize, we get equilibria for parameters, conserved quantities and concentrations parametrized over $[K]$, $K_{\rm tot}>[K]$, $b_2$, and the $a_i$ by the following:
\begin{eqnarray} \label{equilibriapgs}
\overline{KKK}&=&\sqrt{\frac{a_2a_4}{a_1a_3}}\\
KK_{\rm tot}&=&\frac{a_6(K_{\rm tot}-[K])(2+q_1+b_2[K])}{a_5[K](1+b_2(K_{\rm tot}-[K]))}\\
\  [KK] \ &=&\frac{a_6(K_{\rm tot}-[K])}{a_5[K](1+b_2(K_{\rm tot}-[K]))}\\
\overline{KK}&=& \frac{a_6(K_{\rm tot}-[K])(1+b_2[K])}{a_5[K](1+b_2(K_{\rm tot}-[K]))},
\end{eqnarray}
where $q_1=\sqrt{\frac{a_1a_4}{a_2a_3}}$.\\
In order to prove the existence of a Hopf bifurcation it is tempting to proceed
as in the analysis of the MM system for the truncated MAP kinase cascade. To
do so we need to find parameters such that the coefficients $A_0$, $A_1$ and
$A_2$ of the characteristic polynomial of the linearization are all positive.
Then the parameters $a_i, i\in\{1,2,3,4\}$ should be rescaled with a
positive constant $L$ so as to find a value $L_0$ for which the Hurwitz
quantity $H_3=A_1A_2-A_0$ becomes zero. The linearization is very similar to
that for the truncated MAP kinase cascade. Here again one entry is zero and
all others have a sign. After a suitable change in the order of the variables
only one sign differs. The resulting matrix of signs is
\begin{equation}
\left[
{\begin{array}{ccc}
- & 0 & - \\
+ & - & - \\
+ & - & -
\end{array}}
\right].
\end{equation}

The main problem we have encountered in trying to implement this strategy is
to find a point in parameter space where the positivity of the $A_i$ holds.
The coefficient $A_0$ is the negative of the determinant of the linearization.
while the coefficient $A_2$ is the determinant of the analogue of the submatrix
$M_2$ defined in the case of the truncated MAPK cascade (using the modified
order of the variables). Hence the Routh-Hurwitz method will give us the
desired factor $L_0$ only if the determinant of the linearization and that of
$M_2$ have the same (negative) sign. Experiments with Maple indicate that
these two determinants tend to have different signs for parameters satisfying
the biologically motivated positivity conditions. The signs are governed by
the signs of polynomials. When attempting to attain the relevant combination
of signs by fixing some of the parameters and varying others the signs of
the determinants are governed by those of two polynomials. In all experiments
we did these polynomials were different but shared a unique positive zero
where their signs changed. Unfortunately, having the sign of one of the
determinants negative requires being on one side of this zero while having
the other negative requires being on the other side of it. After trying many
parameter configurations we are tempted to to conjecture that there is a
deep reason why these determinants have systematically different signs.
When we apply the Routh-Hurwitz method in the case that $A_0$ and $A_2$ have
different signs it is still possible to find a unique positive value $L_0$
of $L$ with $H_3=0$. However instead of leading to the purely imaginary
eigenvalues required for a Hopf bifurcation this leads to two real
eigenvalues with equal magnitude and opposite sign.

\section{Conclusions and outlook}

The main result of this paper is a rigorous analytical proof of the 
existence of periodic solutions of the Huang-Ferrell system modelling the 
MAPK cascade. Their presence had been suggested by numerical and 
heuristic work in \cite{qiao07}. 
It was also proved that solutions of this type exist for cascades consisting
of a layer with one phosphorylation followed by a layer with two
phosphorylations but for the superficially similar case of a layer with
two phosphorylations followed by a layer with one phosphorylation an
attempt to obtain a similar proof ran into difficulties. The first of these
two cases was considered in \cite{qiao07} but to the authors knowledge the second
was not previously investigated in the literature.
The methods used in the proofs are bifurcation theory and 
geometric singular perturbation theory. It should be noted that the heuristic
considerations in \cite{qiao07}, which might in principle have been used
as a basis for proofs, were in the end hardly used at all. It would be 
interesting to know whether this alternative route could also be effective
in this problem. Relevant ideas, involving relaxation oscillations and the 
Conley index, are explained in \cite{angeli04}, \cite{gedeon07} and 
\cite{gedeon10}. 

An important question is to what extent the mathematical results obtained 
here apply to the real biological system. First it should be noted that in 
nature there is not just one MAPK cascade but many. The basic pattern is 
always the same but the details may be different. We now concentrate on the 
most famous example, the Raf-MEK-ERK cascade. This is essentially the example 
originally considered in \cite{huang96}. In that case Raf is replaced by Mos
but the architecture of the system is identical. In this model it is assumed
that the phosphorylations and dephosphorylations are distributive and 
sequential. Huang and Ferrell concentrate on this case but do mention that they
also did simulations for the alternative versions where one or both of the 
kinases act in a distributive way. In \cite{qiao07}, \cite{zumsande10} and the 
bulk of the present paper only the distributive and sequential case is 
considered. It has been found that phosphorylation of ERK by MEK is 
distributive but not sequential \cite{ferrell97}, \cite{burack97} while 
desphosphorylation of ERK has been found to be distributive and sequential 
\cite{zhao01}. On the other hand the phosphorylation of MEK by Raf has been 
found to be processive \cite{alessi94}. See also \cite{schilling09} where it 
is remarked that the distinction between the two mechanisms may not be 
absolute - processive phosphorylation may be thought of as a limiting case of 
distributive phosphorylation where the second step takes place much faster 
than the first. This means that the original Huang-Ferrell model is not 
applicable to the Raf-MEK-ERK cascade. It is also known that there are cases 
in which processive phosphorylation suppresses complicated dynamical behaviour 
(in this case bistability) present when the phosphorylation is distributive 
\cite{conradi05}, \cite{conradi14}. Thus it would be interesting to 
investigate whether there are oscillations in the system obtained by modifying 
the Huang-Ferrell model by making the phosphorylation in the second layer 
processive. 

It is also interesting to know what effect further interactions between 
proteins not included in the Huang-Ferrell model might have on the 
dynamics. It was observed in \cite{legewie07} that binding of Raf to MEK
can influence the dynamics of the MAPK cascade, enhancing bistability.
In real biological systems the MAPK cascade is also embedded in various 
external feedback loops. One well-known example is that ERK has a suppressive 
effect on Raf via the guanine nucleotide exchange factor son of sevenless (SOS).
That the resulting negative feedback could lead to oscillations was 
observed theoretically in \cite{kholodenko00}. Sustained oscillations in the 
MAPK cascade have been observed experimentally in \cite{shankaran09}. They 
have a period of about 15 minutes and have been observed to continue for over 
ten hours. A quantitative comparison with simulations indicates that these
oscillations are not due to sequestration effects intrinsic to the cascade
but to the feedback loop via SOS. Another type of feedback leading to 
oscillations which involves sequestration but is not intrinsic to the 
cascade is discussed in \cite{liu11}. In that case the binding of activated
ERK to a substrate reduces its availability for dephosphorylation. These 
examples make it clear that there are numerous examples of biological 
interest which represent potential applications of the methods developed in
the present paper. 

Another key issue is that of the biological role of complicated dynamical 
features such as bistability, sustained oscillations or chaos in the MAPK
cascade, with or without external feedback. Many different signals pass 
through this cascade and it may be that non-trivial dynamics can be used to
encode information, for instance by frequency modulation of oscillations.
Here it could be useful to compare with other biological systems where
this type of phenomenon is believed to be important, such as the NF$\kappa$B
pathway \cite{agresti09} or calcium signalling \cite{falcke04}. On the other 
hand it could be that complicated dynamical behaviour in the basic MAPK 
cascade is an unwanted side effect and that the feedback loops in which it is 
embedded in biological systems serve to suppress it. Further mathematical 
investigations of systems related to the MAPK cascade could serve to 
understand the cascade itself better in its biological context and could also
produce new insights into the architecture of biochemical systems. It
should also be kept in mind that a better understanding of the dynamics of 
the cascade could be important for medical progress \cite{ventura09}. In
chemotherapy of cancer Raf inhibitors are already in use while MEK inhibitors
have been the subject of extensive clinical trials but have not yet been 
effective. A better theory of the system could help to understand where to
look for appropriate drugs. 

One question which has been left open here is that of the stability of the 
periodic solutions whose existence has been proved. Is it possible to 
develop methods to prove that in some of these models there are parameters
for which the first Lyapunov coefficient is non-zero, which would solve the
stability problem? Is it possible to extend the techniques used here to
prove the existence of fold-Hopf or Hopf-Hopf bifurcations in the 
Huang-Ferrell model and to check the associated genericity conditions
which would give information on chaotic behaviour? Evidently, modelling the
MAPK cascade gives rise to a large number of challenging mathematical 
problems.

\section*{Appendix: geometric singular perturbation\\ 
theory (GSPT)}

In this appendix some results concerning GSPT needed in the paper will be 
collected. In \cite{hell14} a theorem from \cite{fenichel79} was applied but 
in this paper we need a parameter-dependent version of that result which does 
not obviously follow from the theorem. In fact, starting from the basic 
transformations carried out in \cite{fenichel79} and the idea of a slow 
manifold, the statements we need can be proved using standard results from the 
theory of centre manifolds. This will now be explained.

The starting point is a system of equations of the form
\begin{eqnarray}
&&\dot x=f(x,y,\alpha,\epsilon),\\
&&\epsilon\dot y=g(x,y,\alpha,\epsilon),
\end{eqnarray} 
where $x$, $y$ and $\alpha$ belong to open neighbourhoods of the origin in 
$\R^{n_1}$, $\R^{n_2}$ and $\R^k$ respectively and $\epsilon$ belongs to an
interval of the form $[0,\epsilon_0)$. The dot stands for $d/dt$. It will be 
assumed that the functions $f$ and $g$ are smooth and that they can be 
extended to smooth functions in a neighbourhood of $\epsilon=0$. This system 
is now transformed as in \cite{fenichel79} by defining a new time coordinate
by $\tau=t/\epsilon$ and treating $\epsilon$ and $\alpha$ as new unknowns.
The result is
\begin{eqnarray}
&&x'=\epsilon f(x,y,\alpha,\epsilon),\label{fast1}\\
&&y'=g(x,y,\alpha,\epsilon),\label{fast2}\\
&&\alpha'=0,\label{fast3}\\
&&\epsilon'=0,\label{fast4}
\end{eqnarray}
where the prime denotes $d/d\tau$.
Observe now that the solutions of the equation $g(x,y,\alpha,0)=0$ are
stationary solutions of the system (\ref{fast1})-(\ref{fast4}). We assume 
that $g(0,0,0,0)=0$. Suppose that there exists a smooth function $h_0$
such that $g(x,y,\alpha,0)=0$ is equivalent to $y=h_0(x,\alpha)$. The centre
subspace of the stationary point at the origin is of dimension at least
$n_1+k+1$. We now ensure that its dimension is no greater than that by
assuming that all eigenvalues of the linearization of the system at the 
origin other than the zero eigenvalues arising from the manifold of 
stationary solutions already mentioned and that coming from equation
(\ref{fast4}) have non-zero real parts. These will be called the transverse 
eigenvalues. They are the eigenvalues of $D_yg (0,0,0,0)$. For any positive 
integer $l$ the centre manifold theorem (\cite{kuznetsov10}, Theorem 5.1)
implies that there exists a centre manifold $M_c$ of the origin of class $C^l$ 
and of dimension $n_1+k+1$. Another well-known result about centre manifolds 
states that any stationary solutions sufficiently close to a given stationary 
solution must lie on its centre manifold. Thus the solutions of 
$g(x,y,\alpha,0)=0$ all lie on the centre manifold of the origin. Since the 
dimension of the centre subspace is the same for all of these points it 
follows that $M_c$ is also a centre manifold of these neighbouring points. 
$M_c$ is what is called the slow manifold. In a neighbourhood of the origin it 
can be written in the form $y=h(x,\alpha,\epsilon)$ for a $C^l$ function $h$ 
with $h(x,\alpha,0)=h_0(x,\alpha)$. Considering the restrictions of the 
dynamical 
system with the intersections of $M_c$ with the subspaces of constant $\alpha$ 
and $\epsilon$ gives rise to a dynamical system depending in a regular way on
the parameters $\alpha$ and $\epsilon$. Its explicit form (when written in 
terms of the time coordinate $t$) is 
\begin{equation}
\dot x=f(x,h(x,\alpha,\epsilon),\alpha,\epsilon).
\end{equation}  
In this way the singular limit $\epsilon\to 0$ in the original system has 
been reduced to a regular limit. For $\epsilon=0$ it reduces to
\begin{equation}
\dot x=f(x,h_0(x,\alpha),\alpha,0).
\end{equation}

\end{document}